\documentstyle[11pt,IAU207_pasp,twoside,epsf]{article}
\def\edcomment#1{\iffalse\marginpar{\raggedright\sl#1\/}\else\relax\fi}
\reversemarginpar

\begin{document}
\title{A Survey of M31 Globular Clusters using WFPC2 \\
 on board HST}
 \author{R.M. Rich}
\affil{Dept. of Phys. \& Astron., UCLA, Los Angeles, CA 90095-1562, USA}
\author{C.E. Corsi}
\affil{Osservatorio Astronomico, V.le Parco Mellini 84, 00136 Roma, Italy}
\author{M. Bellazzini, L. Federici, C. Cacciari}
\affil{Osservatorio Astronomico, Via Ranzani 1, 40127 Bologna, Italy}
\author{F. Fusi Pecci}
\affil{Osservatorio Astronomico, Poggio dei Pini, 09012 Capoterra, Italy}

\begin{abstract}
We report new, as yet unpublished HST/WFPC2 V, I photometry of 9
globular clusters in M31.  These are part of a total sample of 19
clusters on M31 with well observed CMDs from HST.   The clusters
have the full range of horizontal branch morphologies from blue HB
to red clump, but none are found with an extreme (blue only) horizontal
branch.  Plots of HB type vs [Fe/H] are similar to those of Galactic
clusters, including a hint of second parameter clusters.
Sixteen halo fields
adjacent to the observed clusters included in our images have
been analysed.  The M31 halo metallicity distribution peaks
at [Fe/H]=--0.7 with a tail toward low metallicity, resembling
the halo of NGC 5128; metal rich
giants appear to be centrally concentrated.

\end{abstract}

\section{The globular clusters}

At present, 22 globular clusters have been observed with FOC or WFPC2 on
board HST.  We report here on the V (F555W)  and I (F814W)  yet unpublished
photometry obtained by Rich and collaborators on the 9 globular clusters most
recently observed (GO6671, PI: R.M. Rich).  The aim of the
project is to compare the stellar populations of the M31
and Milky Way cluster systems, and to measure $M_V^{HB}$ vs
[Fe/H] for a sample of clusters at the same distance.
The data was processed using the HST pipeline, and reduced 
using the ROMAFOT package which is optimized for accurate photometry in
crowded fields.

\begin{figure}
%\centerline{
%\psfig{figure=provafig1.eps,width=13.4cm,height=6cm}
%}
\plotfiddle {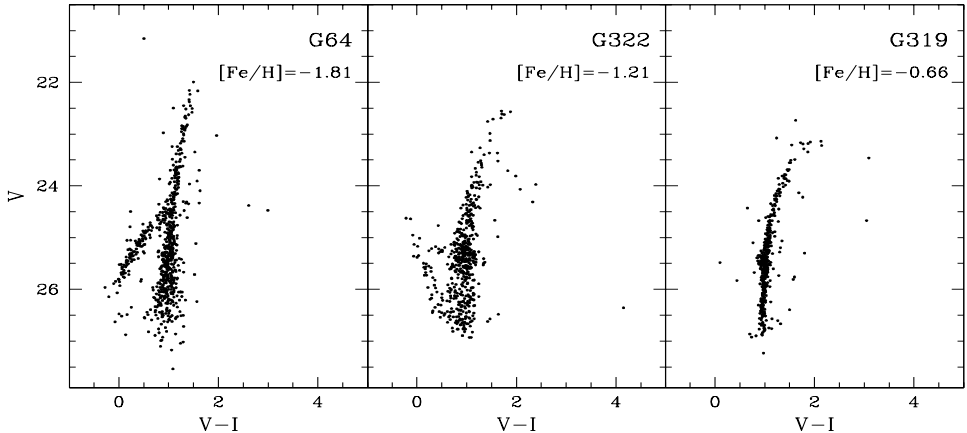} {5cm} {0} {70} {60} {-225pt} {-115pt}
\caption{Romafot Photometry of 3 globular clusters in M31 spanning the full
range in metallicity.  The RGB slope and HB follow the usual
Milky Way trends; G322 may be a second parameter cluster.
The mean accuracy of the data is
$\approx 0.02$ mag in (V, I), $\approx 0.03$ mag in (V-I) for V $<$ 24,
$\approx 0.06$ mag in (V, I), and $\approx 0.08$ mag in (V-I) for V $>$ 24. }
\end{figure}

With the exclusion of 3 clusters from the total sample of 22, because they are
concentrated in the innermost region of the galaxy and their photometry does
not allow the definition of good CMDs, there are now 19 CMDs of sufficiently
good quality to afford some general analysis of the characteristics of the
globular cluster system in M31:

$\bullet$  Our clusters range over $-1.9 <\rm [Fe/H]<-0.6$ with metallicity
from ground-based integrated spectra (Barmby et al. 2000).  The
red giant branch (RGB) morphology and color of
the M31 clusters correlate with metallicity in the
same way as for the Milky Way clusters.

$\bullet$ Horizontal Branch (HB) morphologies range from a
stubby red clump in the most metal-rich clusters to extended blue HBs at
the metal poor end.  We do not find any clusters with extreme blue HBs
(such as M92).  A plot of HB type as a function of [Fe/H] is similar
to that for the Galactic globular clusters, and hints at the existence
of second parameter clusters.

$\bullet$ The dependence of HB luminosity on metallicity is found to be 
$M_V^{HB}=\rm {0.22 [Fe/H]} + const$ where the zero point depends
on the assumed M31 distance.  We find a steeper slope (0.22) than
the value of 0.13 found by Fusi Pecci et al. (1996) based on HST
photometry of 8 M31 globular clusters.

We conclude that the M31 globular clusters show all indications of being
very similar to the Milky Way globular clusters, with no significant anomalies.
There is no strong indication of an intermediate age population of globular
clusters analagous to those found (for example) in the SMC.

\section{Photometry of the M31 halo field population}

\begin{figure}
\plotfiddle {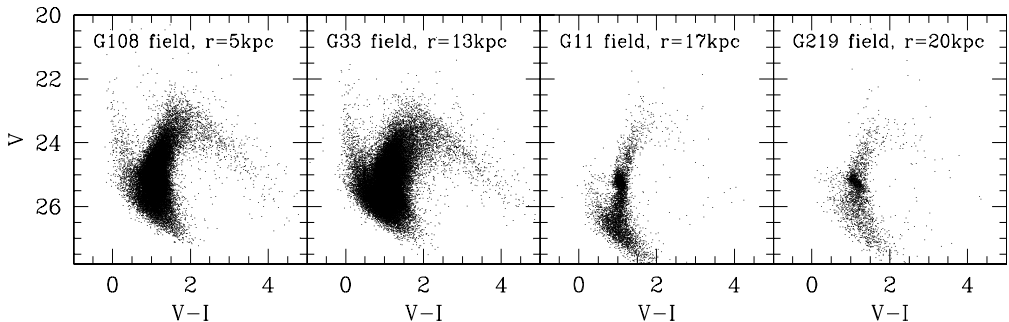} {3cm} {0} {73} {78} {-225pt} {-147pt}
\caption{Color-Magnitude diagrams of 4 fields adjacent to globular clusters
in M31.  Blue sequences are disk stars in M31, while the descending
red giant branch is indicative of high metallicity.  The CMD more
closely resembles the Milky Way bulge than it does a halo field,
even at 20 kpc from the nucleus (G219 field).}
\end{figure}

We have photometry for sixteen fields in the M31 halo, thanks
to the clusters falling on the PC only, making the 3 chips of the
WFC available to image the field population.
These have been reduced in the same way as the clusters,
and a subsample of four of our CMDs is given in Figure 2, in
order of increasing projected galactocentric distance.
The descending red giant branches, indicative of high
metallicity, persist even at 20 kpc from the nucleus.
\begin{figure}
\plotfiddle {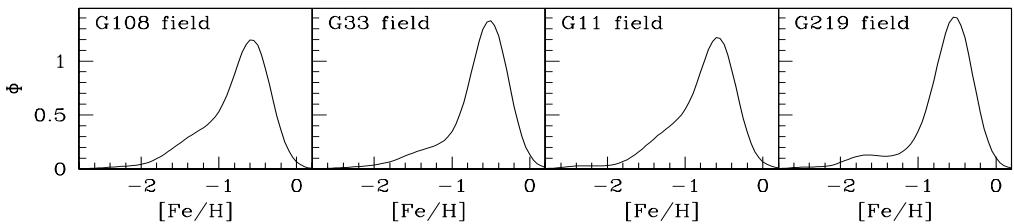} {2.08cm} {0} {73} {63} {-225} {-117}
\caption{Generalized metallicity histograms
 for giants in the 4 halo fields shown in Fig.2; see text.
}
\end{figure}

We estimate metallicities of the field giants by interpolating
their colors between template globular cluster giant branches of
known metallicity.  
We see a strong metal rich peak at $-0.6$ dex in the metallicity 
distribution, with a tail reaching near $-2$ dex (Fig. 3).  Varying
the RGB templates and reddening does not change the fundamental form
of the abundance distribution, which hints
at a second mode near $-1.3$ dex.  We find no
evidence for a gradient in the mean or modal metallicity, but the
fraction of metal rich/metal poor stars increases toward the nucleus.
Our M31 halo metallicity is higher than that of the Milky
Way halo, but the shape and mean agree with
those reported in other galaxies, e.g. NGC5128 (Harris \& Harris, 2000).  It
is difficult to imagine how the accretion of low mass, low metallicity
satellites (such as envisioned by CDM models) could produce such a metal rich halo population. 
Such high metallicities (even 20 kpc from the nucleus)
could be produced by a metal-enriched wind connected with the formation
of the spheroid, with some fraction of halo material coming from
dissolved satellites.

Support for this work was provided by NASA through grant
number GO-6671 from the Space Telescope Science Institute,
which is operated by AURA, Inc., under NASA contract NAS5-26555.


\begin{references}
\reference  Barmby, P., Huchra, J.P., Brodie, J.P., et al. 2000, AJ, 119, 727
\reference Fusi Pecci, F., Buonanno, R., Cacciari, C., et al. 1996, AJ, 112, 1461
\reference Harris, G.L.H., \& Harris, W.E.  2000, AJ, 120, 2423

\end{references}
\end{document}